\documentclass[cits]{PoS}

\title{Decay widths of resonances and pion scattering lengths in a globally
invariant sigma model with vector and axial-vector mesons}

\ShortTitle{A Linear Sigma Model with Vector Mesons and Global Chiral Invariance
}

\author{\speaker{Denis Parganlija}%
\\
        Goethe-Universität Frankfurt am Main\\
        E-mail: \email{parganlija@th.physik.uni-frankfurt.de}}

\author{Francesco Giacosa\\
        Goethe-Universität Frankfurt am Main\\
        E-mail: \email{giacosa@th.physik.uni-frankfurt.de}}

\author{Dirk H. Rischke\\
        Goethe-Universität Frankfurt am Main\\
        E-mail: \email{drischke@th.physik.uni-frankfurt.de}}

\abstract{We calculate low-energy meson decay processes and pion-pion scattering lengths in a two-flavour linear sigma model with global chiral symmetry, exploring the scenario in which the scalar mesons $f_0$(600) and $a_0$(980) are assumed to be $\bar q q$ states.
}

\FullConference{8th Conference Quark Confinement and the Hadron Spectrum \\
		 September 1-6 2008\\
		 Mainz, Germany}

\begin{document}

\section{Introduction}

Effective field theories provide a very efficient means to describe Quantum
Chromodynamics (QCD) at low energies. They possess the same global
symmetries as QCD - e.g., the chiral $SU(N_{f})_{r}\times SU(N_{f})_{l}$
symmetry, where $N_{f}$ is the number of flavours - and are expressed in
terms of hadronic degrees of freedom rather than in terms of quarks and
gluons. Spontaneous breaking of the chiral symmetry leads to the emergence
of low-mass pseudoscalar Goldstone bosons and their chiral partners,
large-mass scalar states.

In this paper we present a linear sigma model with global chiral invariance,
similar to the one of Ref.\ \cite{UBW}. The model contains scalar and
pseudoscalar as well as vector and axialvector mesons. The global invariance
is motivated by the results of Refs. \cite{Reference1, Parganlija:2008xy}
where it has been shown that a locally invariant linear sigma model fails to
describe simultaneously pion-pion scattering lengths and some important
decay widths. For the globally invariant model additional terms appear in
the Lagrangian which introduce new coupling constants that can in principle
be adjusted to improve the agreement with the experimental data. In this
paper we show the first results from this approach for the case of $N_{f}=2$.

As outlined in Ref. \cite{Parganlija:2008xy}, there are two possibilities to
interpret the scalar $\sigma $and $a_{0}$ fields contained in the model
where they are $\bar{q}q$ states $[\sigma =\frac{1}{\sqrt{2}}(\bar{u}u+\bar{d%
}d)$, $a_{0}^{0}=\frac{1}{\sqrt{2}}(\bar{u}u-\bar{d}d)]$: (\textit{i}) they
are identified with $f_{0}(600)$ and $a_{0}(980)$ which form a part of a
larger \ nonet that consists of $f_{0}(980)$, $a_{0}(980)$, $k(800)$ and $%
f_{0}(600)$ (resonances below 1 GeV); (\textit{ii}) they are identified with
the $f_{0}(1370)$ and $a_{0}(1450)$ resonances forming a part of a nonet
that consists of $f_{0}(1370)$, $f_{0}(1500)$, $f_{0}(1710)$, $a_{0}(1450)$, 
$K_{0}(1430)$ - i.e.,\ resonances above 1 GeV (see Ref. \cite{refs1}). In
the assignment (\textit{ii}), scalar mesons below 1 GeV, whose spectroscopic
wave functions possibly contain a dominant tetraquark or mesonic molecular
contribution \cite{refs2}, may be introduced as additional scalar fields.

In this paper, we describe briefly the consequences of assignment \textit{(i)}; the consequences of assumption \textit{(ii)} as well as more detailed
results in assignment \textit{(i)} may be found in Ref. \cite{PGR}.

\section{The Linear Sigma Model with Global Chiral Symmetry}

The Lagrangian of the globally invariant linear sigma model with $%
U(2)_{R}\times U(2)_{L}$ symmetry reads \cite{Reference1, Parganlija:2008xy,
Boguta}: 
\begin{eqnarray}
\mathcal{L} &=&\mathrm{Tr}[(D^{\mu }\Phi )^{\dagger }(D^{\mu }\Phi
)]-m_{0}^{2}\mathrm{Tr}(\Phi ^{\dagger }\Phi )-\lambda _{1}[\mathrm{Tr}(\Phi
^{\dagger }\Phi )]^{2}-\lambda _{2}\mathrm{Tr}(\Phi ^{\dagger }\Phi )^{2} 
\nonumber \\
&&-\frac{1}{4}\mathrm{Tr}[(L^{\mu \nu })^{2}+(R^{\mu \nu })^{2}]+\frac{%
m_{1}^{2}}{2}\mathrm{Tr}[(L^{\mu })^{2}+(R^{\mu })^{2}]+\mathrm{Tr}[H(\Phi
+\Phi ^{\dagger })]  \nonumber \\
&&+c(\det \Phi +\det \Phi ^{\dagger })-2ig_{2}(\mathrm{Tr}\{L_{\mu \nu
}[L^{\mu },L^{\nu }]\}+\mathrm{Tr}\{R_{\mu \nu }[R^{\mu },R^{\nu }]\}) 
\nonumber \\
&&-2g_{3}\{\mathrm{Tr}[(\partial _{\mu }L_{\nu }+\partial _{\nu }L_{\mu
})\{L^{\mu },L^{\nu }\}]+\mathrm{Tr}[(\partial _{\mu }R_{\nu }+\partial
_{\nu }R_{\mu })\{R^{\mu },R^{\nu }\}]\}+\mathcal{L}_{4}  \nonumber \\
&&+\frac{h_{1}}{2}\mathrm{Tr}(\Phi ^{\dagger }\Phi )\mathrm{Tr}[(L^{\mu
})^{2}+(R^{\mu })^{2}]+h_{2}\mathrm{Tr}[(\Phi R^{\mu })^{2}+(L^{\mu }\Phi
)^{2}]+2h_{3}\mathrm{Tr}(\Phi R_{\mu }\Phi ^{\dagger }L^{\mu }),
\label{Lagrangian}
\end{eqnarray}%
with $\Phi =(\sigma +i\eta _{N})\,t^{0}+(\vec{a}_{0}+i\vec{\pi})\cdot \vec{t}
$ (scalar and pseudoscalar mesons; our model is valid for $N_{f}=2$ and thus
our eta meson $\eta _{N}$ contains only non-strange degrees of freedom); $%
L^{\mu }=(\omega ^{\mu }-f_{1}^{\mu })\,t^{0}+(\vec{\rho}^{\mu }-\vec{a}%
_{1}^{\mu })\cdot \vec{t}$ and $R^{\mu }=(\omega ^{\mu }+f_{1}^{\mu
})\,t^{0}+(\vec{\rho}^{\mu }+\vec{a}_{1}^{\mu })\cdot \vec{t}$ (vector and
axialvector mesons), where $t^{0}$, $\vec{t}$ are the generators of $U(2)$; $%
D^{\mu }\Phi =\partial ^{\mu }\Phi +ig_{1}(\Phi L^{\mu }-R^{\mu }\Phi )$, $%
L^{\mu \nu }=\partial ^{\mu }L^{\nu }-\partial ^{\nu }L^{\mu }$, $R^{\mu \nu
}=\partial ^{\mu }R^{\nu }-\partial ^{\nu }R^{\mu }$ and $\mathcal{L}_{4}$
containing all the vertices with four vector and axialvector mesons.
Explicit breaking of the global symmetry is described by the term Tr$[H(\Phi
+\Phi ^{\dagger })]\equiv h\sigma (h=const.)$. The chiral anomaly is
described by the term $c\,(\det \Phi +\det \Phi ^{\dagger })$ \cite{Hooft}.

Irrespective of $\mathcal{L}_{4}$,\ the model contains 13 parameters - 12
parameters from the Lagrangian (\ref{Lagrangian}), plus the wave function
renormalisation constant of the pseudoscalar mesons \cite{RS}, Z. However,
only seven of those ($Z$, $g_{1,2}$, $h_{1,2,3}$, $\lambda _{2}$) are
relevant for the decays that will be considered in the following. The
parameters $g_{1}$, $h_{3}$ and $\lambda _{2}$\ are expressed in terms of $Z$%
:

\[
g_{1}=g_{1}(Z)=\frac{m_{a_{1}}}{Zf_{\pi }}\sqrt{1-\frac{1}{Z^{2}}}%
,\;\;h_{3}=h_{3}(Z)=\frac{1}{Z^{2}f_{\pi }^{2}}\left( m_{\rho }^{2}-\frac{%
m_{a_{1}}^{2}}{Z^{2}}\right) ,\;\;\lambda _{2}=\lambda _{2}(Z)=\frac{%
Z^{2}m_{a_{0}}^{2}-m_{\eta _{N}}^{2}}{Z^{4}f_{\pi }^{2}}, 
\]%
and thus the number of independent relevant parameters is decreased to four.
Additionally, $m_{\sigma }$ (which is a function of $m_{0}$, $\lambda _{1}$, 
$\lambda _{2}$, $c$ and $Z$) is taken as a parameter that can be determined
from the pion-pion scattering lengths yielding five independent parameters
for the meson decay modes and scattering lengths described below.

\subsection{Relevant Decay Modes and $\pi \pi$ Scattering Lengths}

The following decay modes of two-flavour low-energy mesons have been taken
into account [parameter dependence in brackets]: $\rho \rightarrow \pi \pi $ 
$[Z$, $g_{2}]$, $f_{1}\rightarrow a_{0}\pi $ $[Z$, $h_{2}]$, $%
a_{1}\rightarrow \pi \gamma $ $[Z]$, $a_{0}\rightarrow \eta \pi $ $[Z$, $%
h_{2}]$, $\sigma \rightarrow \pi \pi $ $[Z$, $h_{1}$, $h_{2}]$, $%
a_{1}\rightarrow \sigma \pi $ $[h_{1}$, $h_{2}$, $Z]$, $a_{1}\rightarrow
\rho \pi $ $[g_{2}$, $Z]$. We have also considered the pion-pion scattering
lengths $a_{0}^{0}(h_{1}$, $h_{2}$, $Z$, $m_{\sigma })$ and $a_{0}^{2}(h_{1}$%
, $h_{2}$, $Z$, $m_{\sigma })$.

Given that the decay widths for the channels $\sigma \rightarrow \pi \pi $ $%
[Z$, $h_{1}$, $h_{2}]$, $a_{1}\rightarrow \sigma \pi $ $[h_{1}$, $h_{2}$, $%
Z] $ and $a_{1}\rightarrow \rho \pi $ $[g_{2}$, $Z]$ are poorly known, we
have not taken any numerical values for these decay widths to fit our
parameters - these decay widths will be calculated as a consistency check on
the basis of the results obtained from the other decay widths and the
scattering lengths.

Here, we will present formulas that have been used to fit the parameters;
for all other formulas, see Ref. \cite{PGR}. \newline

\textit{Decay width for} $\rho \rightarrow \pi \pi $\textit{.} The decay
width reads

\[
\Gamma _{\rho \rightarrow \pi \pi }=\frac{m_{\rho }^{5}}{48\pi m_{a_{1}}^{4}}%
\left[ 1-\left( \frac{2m_{\pi }}{m_{\rho }}\right) ^{2}\right] ^{\frac{3}{2}}%
\left[ \left( g_{1}-\frac{g_{2}}{2}\right) Z^{2}+\frac{g_{2}}{2}\right]
^{2}. 
\]

The experimental value is (149.4$\pm 1.0$) MeV \cite{PDG}. \newline

\textit{Decay width for} $f_{1}\rightarrow a_{0}\pi $\textit{.} The
following formula for the decay width is obtained:

\[
\Gamma _{f_{1}\rightarrow a_{0}\pi }=\frac{g_{1}^{2}Z^{2}}{2\pi }\frac{k^{3}%
}{m_{f_{1}}^{2}m_{a_{1}}^{4}}\left[ m_{\rho }^{2}-\frac{1}{2}%
(h_{2}+h_{3})\phi ^{2}\right] ^{2},\,k=\frac{1}{2}\sqrt{m_{f_{1}}^{2}-2(m_{%
\pi }^{2}+m_{a_{0}}^{2})+\frac{(m_{a_{0}}^{2}-m_{\pi }^{2})^{2}}{%
m_{f_{1}}^{2}}} 
\]

where $\phi \equiv Zf_{\pi }$ is the vacuum expectation value of the $\sigma 
$ field. \newline

\textit{Decay width for} $a_{1}\rightarrow \pi \gamma $\textit{. }The
Lagrangian leading to the formula for the decay width $\Gamma
_{a_{1}\rightarrow \pi \gamma }$ is obtained from the Lagrangian (\ref%
{Lagrangian}) by coupling the photon to the relevant part of the axial
current $J_{\mu }=-ig_{1}Z^{2}f_{\pi }(a_{1\mu }^{+}\pi ^{-}-a_{1\mu
}^{-}\pi ^{+})-Zw(a_{1\mu \nu }^{+}\partial ^{\nu }\pi ^{-}-a_{1\mu \nu
}^{-}\partial ^{\nu }\pi ^{+})$, where $w=\frac{g_{1}\phi }{m_{a_{1}}^{2}}$,
and reads $\mathcal{L}_{a_{1}\pi \gamma }=eJ_{\mu }A^{\mu }$.

The decay width reads

\[
\Gamma _{a_{1}\rightarrow \pi \gamma }=\frac{e^{2}}{96\pi }(Z^{2}-1)m_{a_{1}}%
\left[ 1-\left( \frac{m_{\pi }}{m_{a_{1}}}\right) ^{2}\right] ^{3}. 
\]

Note that the sole dependence of the $a_{1}\rightarrow \pi \gamma $ decay
width on the parameter $Z$ may in principle lead to an accurate
determination of this parameter. However, the experimental value of the $%
a_{1}\rightarrow \pi \gamma $ decay width is not very precise ($\Gamma
_{a_{1}\rightarrow \pi \gamma }=0.640\pm 0.246$ MeV \cite{PDG}) and thus we
have used the $\chi ^{2}$ method to determine all the parameters from the
decay widths and scattering lengths. \newline

\textit{Decay amplitude for} $a_{0}\rightarrow \eta _{N}\pi $\textit{.} The
mass of the $\eta _{N}$ meson can be calculated using the well-known mixing
of strange and non-strange contributions in the physical fields $\eta $ and $%
\eta ^{\prime }(958)$ yielding $\eta =\eta _{N}\cos \varphi +\eta _{S}\sin
\varphi $; $\eta ^{\prime }=-\eta _{N}\sin \varphi +\eta _{S}\cos \varphi $, 
where $\eta _{S}$ denotes a pure $\bar{s}s$ state and $\varphi \simeq -36%
{{}^\circ}%
$ \cite{Giacosa:2007up}. Then we obtain $m_{\eta }=716$ MeV.

Note that we have used the decay amplitude for the $a_{0}\rightarrow \eta
_{N}\pi $ decay instead of the decay width as quoted by the PDG \cite{PDG}
in order to fit the parameters of the model. The experimental value of the
decay amplitude is known from the Crystal Barrel data: $A_{a_{0}\eta \pi
}=(3330\pm 150)$ MeV \cite{Bugg} which for our purposes has to be divided by 
$\cos \varphi $; the formula for the decay amplitude obtained from Eq. (\ref%
{Lagrangian}) is

\[
A_{a_{0}\eta \pi }=\frac{m_{\eta }^{2}-Z^{2}m_{a_{0}}^{2}}{\phi }+\frac{%
g_{1}^{2}\phi }{m_{a_{1}}^{2}}\left\{ \left[ 1-\frac{1}{2}\frac{Z^{2}\phi
^{2}}{m_{a_{1}}^{2}}(h_{2}-h_{3})\right] (m_{a_{0}}^{2}-m_{\pi }^{2}-m_{\eta
}^{2})+Z^{2}m_{a_{0}}^{2}\right\} . 
\]

\textit{Scattering length} $a_{0}^{0}$\textit{.} The formula for $a_{0}^{0}$
is calculated using the partial wave decomposition \cite{ACGL} which leads to

\begin{eqnarray*}
\lefteqn{
a_{0}^{0}=\frac{1}{4\pi } \left \{2g_{1}^{2}Z^{4}\frac{m_{\pi }^{2}}{%
m_{a_{1}}^{4}}\left[ m_{\rho }^{2}+\frac{\phi ^{2}}{16}%
(12g_{1}^{2}-2(h_{1}+h_{2})-14h_{3})\right] -\frac{5}{8}\frac{Z^{2}m_{\sigma
}^{2}-m_{\pi }^{2}}{f_{\pi }^{2}} \right.
}
 \\
&&
-\frac{3}{2}\left[ 2 g_{1}^{2}Z^{2}\phi \frac{m_{\pi }^{2}}{m_{a_{1}}^{2}}
\left( 1+\frac{m_{\rho }^{2}-\phi ^{2}( h_{1}+h_{2}+h_{3}) /2}{2
m_{a_{1}}^{2}} \right )- \frac{Z^{2}m_{\sigma }^{2}-m_{\pi }^{2}}{2 \phi }%
\right] ^{2}\frac{1}{4m_{\pi }^{2}-m_{\sigma }^{2}} \\
&&
+ \left. \left[ g_{1}^{2}Z^{2}\phi \frac{m_{\pi }^{2}}{m_{a_{1}}^{4}}\left(
m_{\rho }^{2}-\frac{\phi ^{2}}{2}(h_{1}+h_{2}+h_{3})\right) +
\frac{Z^{2}m_{\sigma }^{2}-m_{\pi }^{2}}{2 \phi }\right] ^{2}\frac{1}{%
m_{\sigma }^{2}} \right \}.
\end{eqnarray*}

We are using the result $a_{0}^{0}=0.233\pm 0.023$ (normalised to the pion
mass) in accordance with data published by the NA48/2 collaboration \cite{SL}%
. \newline

\textit{Scattering length }$a_{0}^{2}$\textit{.} An analogous calculation as
in the case of the scattering length $a_{0}^{0}$ leads to

\begin{eqnarray*}
a_{0}^{2} &=&-\frac{1}{4\pi } \left \{\frac{1}{4}\frac{Z^{2}m_{\sigma }^{2}-m_{\pi
}^{2}}{f_{\pi }^{2}}+g_{1}^{2}Z^{4}\frac{m_{\pi }^{2}}{m_{a_{1}}^{4}}\left[
m_{\rho }^{2}-\frac{\phi ^{2}}{2}(h_{1}+h_{2}+h_{3})\right] \right. \\
&&- \left. \left[ g_{1}^{2}Z^{2}\phi \frac{m_{\pi }^{2}}{m_{a_{1}}^{4}}\left(
m_{\rho }^{2}-\frac{\phi ^{2}}{2}(h_{1}+h_{2}+h_{3})\right) +\frac{%
Z^{2}m_{\sigma }^{2}-m_{\pi }^{2}}{2\phi }\right] ^{2}\frac{1}{m_{\sigma
}^{2}} \right \}.
\end{eqnarray*}

The result for $a_{0}^{2}$ from the NA48/2 collaboration \cite{SL} is $%
a_{0}^{2}=-0.0471\pm 0.015$.

\section{Results}

In order to fit the relevant parameters of our model ($Z$, $g_{2}$, $h_{1,2}$%
, $m_{\sigma }$) to experimental data (for the aforementioned decay widths, $%
a_{0}\rightarrow \eta \pi $ decay amplitude and scattering lengths) we have
used the $\chi ^{2}$ method. 
The error for the mixing angle $\varphi $ is neglected in this first case
study.\newline
Our best fit yields the minimal value of $\chi _{\mathrm{min.}}^{2}=0.752516$
per degree of freedom which leads to the following values of parameters: $%
Z=1.5217$, $g_{1}=6.59$, $g_{2}=0.3365$, $h_{1}=-100.7$, $h_{2}=106.045$, $%
h_{3}=-2.63$, $m_{\sigma }=330$ MeV. 
\newline
It is interesting to note that, although new parameters have been introduced
in the globally invariant model, the values of $Z=1.5217$, $g_{1}=6.59$, and 
$m_{\sigma }=330$ MeV are virtually the same as those obtained in the
locally invariant model where the corresponding values were $Z=1.586$, $%
g_{1}=6.51$, and $m_{\sigma }\simeq (315-345)$ MeV \cite{Parganlija:2008xy}.
Note also that the value of $h_{1}$ does not appear to be large-$N_{C}$
suppressed, although the parameter $h_{1}/2$ is the prefactor to a term
consisting of a product of two traces Tr$(\Phi ^{\dagger }\Phi )$Tr$[(L^{\mu
})^{2}+(R^{\mu })^{2}]$ - in fact, the modulus of the corresponding
prefactor $h_{1}/2$ is by about a factor of ten larger than the prefactor to
the term Tr$(\Phi R_{\mu }\Phi ^{\dagger }L^{\mu })$ (i.e., $2h_{3}=-5.26$).

Using the parameters above leads to the following consequences: (\textit{i}) 
$\Gamma _{a_{1}\rightarrow \sigma \pi }=90.163$ MeV; (\textit{ii}) given
that in the globally invariant model the $\rho $ mass term is $m_{\rho
}^{2}=m_{1}^{2}+\phi ^{2}(h_{1}+h_{2}+h_{3})/2$, it is possible to calculate
the contribution of the bare mass ($m_{1}^{2}$) to the total mass $m_{\rho
}^{2}$\ - the result $m_{1}\simeq $ 758 MeV is obtained, leading to a very
small contribution of the quark condensate to the $\rho $ mass; (\textit{iii}%
) the $\sigma \rightarrow \pi \pi $\ decay width has a value of less than 10
MeV - it is thus too small - and the $a_{1}\rightarrow \rho \pi $ decay
width has the value of 1.4 GeV - it is thus too large.

Hence, in the light of our results we conclude that the $\bar{q}q$
assignment of the light scalar mesons leads to contradictions to experiment.
For a definite conclusion, the errors of the parameters in the model should
be evaluated (see Ref. \cite{PGR}), but it is already clear from our current
results that the assignment of $f_{0}(600)$ and $a_{0}(980)$ as $\bar{q}q$
states may be problematic.

A possible way to resolve the aforementioned problem is to redefine $\sigma $
and $a_{0}$ mesons in the model as $f_{0}(1370)$ and $a_{0}(1450)$,
respectively, and hence assign the scalar meson states to the energy region
above 1 GeV \cite{PGR}. Then, the mixing of quarkonia and tetraquark states 
\cite{Heinz} needs to be examined.

\section{Conclusions and Outlook}

A globally invariant linear sigma model with vector and axial vector mesons
and its consequences for low-energy meson decay channels and pion-pion
scattering lengths have been presented. Results obtained in the assignment
in which scalar mesons are identified as states under 1 GeV indicate
contradictions to experimental data, hence raising questions about the
justification of the mentioned assignment. Thus, a detailed study of the
other possible assignment for scalar mesons (in which those states are
located in the energy region above 1 GeV) is necessary. In the future, other
relevant issues in connection with vacuum phenomenology will be addressed
such as the inclusion of the nucleon field together with its chiral partner 
\cite{Susanna} as well as extending the work of Refs. \cite{RS,Heinz} to
consider chiral symmetry restoration at nonzero temperature.

\acknowledgments

The authors thank H. Leutwyler and S. Str\"{u}ber for valuable discussions
during the preparation of this work.

\end{document}